\documentclass{article}

\usepackage[utf8]{inputenc}
\usepackage[T1]{fontenc}
\usepackage[english]{babel}
\usepackage{amsmath}
\usepackage{amssymb}
\usepackage{float}
\usepackage{graphicx}

\title{Recursive double-size fixed precision arithmetic}
\author{C.Chabot, JG. Dumas, L. Fousse, P. Giorgi\\
\{christophe.chabot,jean-guillaume.dumas,laurent.fousse\}@imag.fr\\ pascal.giorgi@lirmm.fr}

\begin{document}

\maketitle

\section{Introduction}
This work is a part of the SHIVA (Secured Hardware Immune Versatile Architecture) project whose purpose is to provide a programmable and reconfigurable hardware module with high level of security. We propose a recursive double-size fixed precision arithmetic called RecInt. Our work can be split in two parts. First we developped a C++ software library with performances comparable to GMP ones. Secondly our simple representation of the integers allows an implementation on FPGA. \\

Concerning the software part, we remarked that most often the general purpose arbitrary precision GMP library is faster for cryptographic routines than special purpose libraries such as OpenSSL or Miracl. Then we found that GMP could be improved on very small precision. Our idea is to consider sizes that are a power of 2 and to apply doubling techniques to implement them efficiently: we design a recursive data structure where integers of size $2^k$, for $k > k_0$ can be stored as two integers of size $2^{k-1}$. Obviously for $k\leq k_0$ we use machine arithmetic instead ($k_0$ depending on the architecture). Our design makes use of C++ template mechanism so that we can define a generic doubling structure for large $k$ and specialize it to machine arithmetic for small values of $k$. If some routines can be implemented faster for some specific sizes, the template mechanism allows also partial specializations of these routines. \\
We provide a prototype implementation showing good performances on desktop PC's: the PALOALTO library\footnote{https://www.ljkforge.imag.fr/projects/paloalto/.}. \\

Concerning the hardware part, our first works are based on the transformation of C++ sources to VHDL using dedicated softwares. We show that our first results are promising. \\

\section{Recursive data structure}

The main idea is to represent an element of size $2^k$ with two elements of size $2^{k-1}$. We note \verb?RecInt<?$k$\verb?>? our recursive integer of size $2^k$ bits. That leads to a recursive structure with specializations for small sizes. 
$$\left\{\begin{array}{l}\verb?RecInt<?k\verb?>? \  a\mapsto \ (\verb?RecInt<? k-1 \verb?>?\ a.High,\verb? RecInt<? k-1 \verb?>?\ a.Low) \\
\verb?RecInt<?k_0\verb?>?\text{ of size }2^{k_0} \text{ called \textbf{limb}}.
\end{array}\right.$$

$$\text{where }\quad a=a.High*2^{2^{k-1}}+a.Low.$$

All these integers are unsigned, hence a \verb?RecInt<?$k$\verb?>? is capable to store any unsigned integer within the range $0..2^{2^k}$. \\

\begin{figure}[H]
  \begin{center}
    \includegraphics[width=300px]{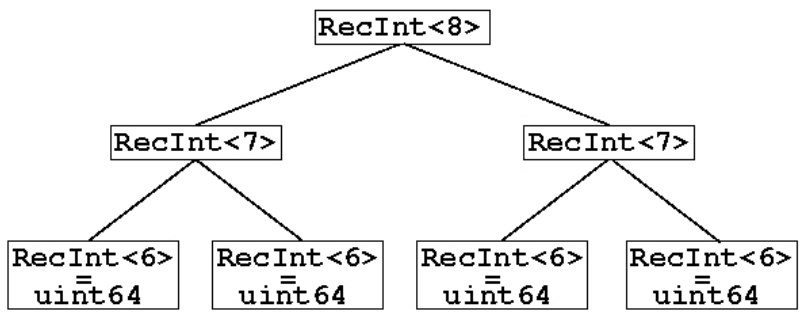} \\
    \ \\
    \caption{Recursive structure of a RecInt<8> in a 64 bits architecture.}
  \end{center}
\end{figure}

\section{Operations}

Obviously all the classical arithmetic operations are provided. However we focus on specific ones.

\subsection{Extending the word size}

The idea is to provide a very fast arithmetic for integers of size a power of two which would mimic the behaviour of the word-size arithmetic: operations are correct modulo some $2^{2^k}$. Indeed computing the remainder with such a modulus on a binary architecture comes to just keeping the correct number of bits. \\
For instance, machine size arithmetic on a 32 bits architecture is done modulo $2^{2^5}$, and modulo $2^{2^6}$ on a 64 bits architecture. \\

For truncated addition and subtraction, that comes to not keeping the carry (or the borrow). \\

When multiplying two integers of at most $2^k$ bits modulo $2^{2^k}$, one does not need to compute the highest bits of the product as in the following figure. \\

\begin{figure}[H]
  \begin{center}
    \includegraphics[width=300px]{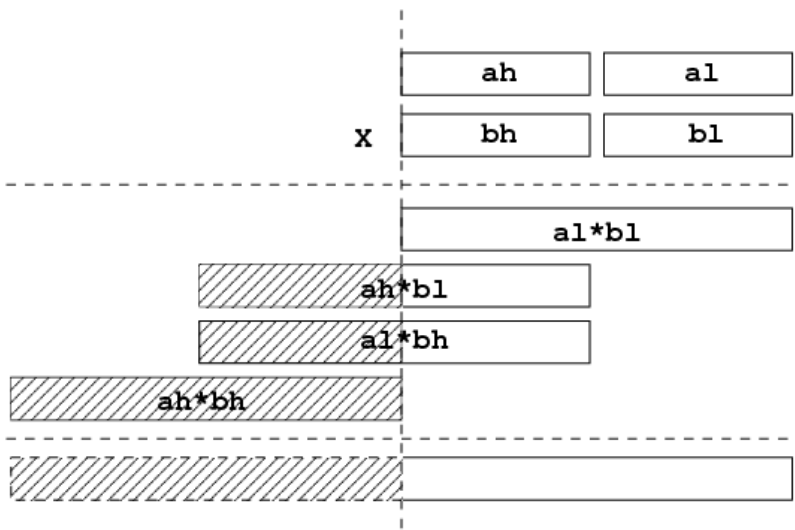}
    \caption{Truncated multiplication}
  \end{center}
\end{figure}

Hence, one truncated multiplication of level $k$ requires only 1 complete multiplication and 2 truncated multiplications of level $k-1$ (instead of $4$ complete multiplications for a naïve complete multiplication). \\

\subsection{Recursive division}
We use a recursive method for Euclidean division of integers described in \cite{Burn}. This method uses two sub-algorithms dividing respectively 2 digits by 1 digit and 3 halves by 2. They allow then a recursive division of a $s$-digits integer by a $r$-digits integer with complexity $O(rs^{\log(3)-1}+r\log(s))$.

\subsection{Montgomery modular multiplication}
A naïve way of performing a modular multiplication is: performing a complete multiplication and then a modular reduction. However in the general case, this reduction is done with a division, which is time consuming. In the case there exists a radix $R$ such that computations modulo $R$ are inexpensive to process, Montgomery gives in \cite{Mont} a method for performing a modular multiplication without trial division. Actually the complete multiplication is still performed but the reduction is done as following: \\
\newpage
\noindent\textbf{function} REDC($T$) \\
\hspace*{0.5cm}$m\leftarrow (T\text{ mod }R)N'\text{ mod }R$\\
\hspace*{0.5cm}$t\leftarrow (T+mN)/R$\\
\hspace*{0.5cm}\textbf{if }$t\geq N$\textbf{ then return }$t-N$\textbf{ else return }$t$. \\

where $0<R^{-1}<N$ and $O<N'<R$ satisfy $RR^{-1}-NN'=1$.\\

This algorithm computes REDC($T$) $=TR^{-1}$ mod $N$ if $0\leq T < RN$. \\
Thus, REDC($(AR\text{ mod }N)(BR\text{ mod }N)$)$=ABR\text{ mod }N$. This means that if we note $\overline{A}=AR\text{ mod }N$ and $\overline{B}=BR\text{ mod }N$ (called Montgomery representations of $A$ and $B$), then REDC($\bar{A}\bar{B}$)=$\overline{AB}$. \\

Hence, as long as we stay in Montgomery representation no division is required. In order to come back to the regular representation, one needs to perform a multiplication by $R^{-1}$ modulo $N$. \\

In our case, if one wants to multiply two \verb?RecInt<?$k$\verb?>? $A$ and $B$, $R=2^{2^k}$. If $A$ is a \verb?RecInt<?$k+1$\verb?>?,
\begin{itemize}
\item Reduction of $A$ modulo $R$ comes to take $A.Low$.
\item Exact division of $A$ by $R$ comes to take $A.High$. \\
\end{itemize} 

Hence REDC requires only 1 truncated multiplication and 1 complete multiplication.

\section{C++ Library}

This library depends on \verb?gmp? (for machine word arithmetic). Hence if \verb?gmp? is intalled on your computer you can use our library by typing at the beginning of your C++ program:
\begin{verbatim}
#define __32bits  // or __64bits depending on your processor
#include "recint.h"
\end{verbatim}

\subsection{Template recursive data structure}

In order to simplify the development we chose to use a template recursive data structure with partial specialization. This specialization depends on the architecture. 
\begin{itemize}
\item \verb?RecInt<5>? $\sim$ \verb?uint32? in a 32 bits architecture.
\item \verb?RecInt<6>? $\sim$ \verb?uint64? in a 64 bits architecture.
\end{itemize}

First our library allows manipulation of fixed precision integers. Thus we define the template structure \verb?RecInt<>? as following: \\

\begin{verbatim}
template <size_t k> struct RecInt {
  typedef RecInt<k+1> Father_t;
  typedef RecInt<k> Self_t;
  typedef RecInt<k-1> Half_t;

  // High = most significant part
  // Low = least significant part
  // *this == High * 2^(2^(k-1)) + Low
  Half_t High, Low;
};

template <> struct RecInt<LIMB_SIZE> {
  limb Value;
};
\end{verbatim}
where \verb?limb? represents the machine word. \\

\subsection{Classical operations}

Parameters of functions have been chosen to be passed by reference in order to avoid copying them at each call. \\

Functions that return a boolean are always of the following form:
\begin{verbatim}
template <class T> bool RI_is_equal_to(const T&, const T&);
\end{verbatim}
All the other functions are \verb?void? functions whose first parameters are the output values and the last ones, considered as \verb?const? are the input ones:
\begin{verbatim}
template <class T> void RI_add(limb&, T&, const T&, const T&);
\end{verbatim}

Classic functions are split into the following sections according to their use:

\subsubsection{Operators}

We define the following operators for basic arithmetic.
\begin{itemize}
\item Arithmetic operators: \verb?+?, \verb?-?, \verb?*?;
\item In place operators: \verb?+=?, \verb?-=?, \verb?*=?;
\item Increment and decrement operators: \verb?++?, \verb?--?;
\item Comparison operators: \verb?==?, \verb?!=?, \verb?<?, \verb?>?;
\end{itemize}

\subsubsection{Comparison functions}

Comparison between \verb?RecInt?:
  \begin{itemize}
  \item \verb?char RI_comp(const RecInt<k>& a, const RecInt<k>& b)? returns \verb?+1? if \verb?a>b?, \verb?0? if \verb?a==b? and \verb?-1? if \verb?a<b?.
  \item \verb?bool RI_is_equal_to(const RecInt<k>& a, const RecInt<k>& b)? returns \verb?true? if and only if \verb?a==b?.
  \item \verb?bool RI_is_greater_than(const RecInt<k>& a, const RecInt<k>& b)? returns \verb?true? if and only if \verb?a>b?.
  \item \verb?bool RI_is_lower_than(const RecInt<k>& a, const RecInt<k>& b)? returns \verb?true? if and only if \verb?a<b?.
  \end{itemize}
  Comparison with a constant:
  \begin{itemize}
  \item \verb?bool RI_is_equal_to_0(const RecInt<k>& a)? returns \verb?true? if and only if \verb?a==0?.
  \item \verb?bool RI_is_equal_to_1(const RecInt<k>& a)? returns \verb?true? if and only if \verb?a==1?.
  \item \verb?bool RI_is_equal_to_limb(const RecInt<k>& a, const limb& b)? returns \verb?true? if and only if \verb?a==b?.
  \end{itemize}

\subsubsection{Set functions}
We provide a set of functions permitting to set a part of an integer or the whole integer to a specified value.
\begin{itemize}
\item \verb?void RI_reset(RecInt<k>& a)? resets \verb?a? to 0.
\item \verb?void RI_random(RecInt<k>& a)? sets \verb?a? to a random value.
\item \verb?void RI_set_limb(RecInt<k>& a, const limb& b, const unsigned int& n)? sets the $n^{th}$ limb of \verb?a? to the value \verb?b? (the $0^{th}$ limb is the least significant one).
\item \verb?void RI_set_const(RecInt<k>& a, const limb& b)? sets the $0^{th}$ limb of \verb?a? to the value \verb?b?.
  
\end{itemize}

\subsubsection{Get functions}
The following functions permit to get value(s) from an integer:
\begin{itemize}
\item \verb?void RI_get_limb(limb& l, const RecInt<k>& a, const unsigned int& n)?: \verb?l? is set to the value of the $n^{th}$ limb of \verb?a?.
\item \verb?void RI_get_limb0(limb& l, const RecInt<k>& a)?: \verb?l? is set to the value of the least significant limb of \verb?a?.
\item \verb?void RI_get_limbn(limb& l, const RecInt<k>& a)?: \verb?l? is set to the value of the most significant limb of \verb?a?.
\item \verb?void RI_copy(RecInt<k>& a, const RecInt<k>& b)? copies the value of \verb?b? into \verb?a?.  
\end{itemize}

\subsection{Arithmetic functions}

The following classic operations will always output the full precision of the computation. 

\begin{itemize}
\item \verb?void RI_add(limb& l, RecInt<k>& a, const RecInt<k>& b , const RecInt<k>& c)?: $a=b+c+l*2^{2^k}$.
\item \verb?void RI_add(limb& l, RecInt<k>& a, const RecInt<k>& b , const limb& c)?: $a=b+c+l*2^{2^k}$.
\item \verb?void RI_increment(limb& l, RecInt<k>& a)?: $a\leftarrow a+1+l*2^{2^k}$.
\item \verb?void RI_sub(limb& l, RecInt<k>& a, const RecInt<k>& b , const RecInt<k>& c)?: $a=b-c+l*2^{2^k}$.
\item \verb?void RI_sub(limb& l, RecInt<k>& a, const RecInt<k>& b , const limb& c)?: $a=b-c+l*2^{2^k}$.
\item \verb?void RI_decrement(limb& l, RecInt<k>& a)?: $a\leftarrow a-1+l*2^{2^k}$.
\item \verb?void RI_lmul(RecInt<k>& ah, RecInt<k>& al, const RecInt<k>& b , const RecInt<k>& c)?: $ah*2^{2^k}+al=b*c$.
\item \verb?void RI_lmul(limb& ah, RecInt<k>& al, const RecInt<k>& b , const limb& c)?: $ah*2^{2^k}+al=b*c$.
\item \verb?void RI_div(RecInt<k>& q, RecInt<k>& r, const RecInt<k>& a , const RecInt<k>& b)?: \verb?q? and \verb?r? are respectively the quotient and the remainder in the Euclidean division of \verb?a? by \verb?b?: $a=b*q+r$ with $r<b$.
\item \verb?void RI_div_quotient(RecInt<k>& q, const RecInt<k>& a , const RecInt<k>& b)?: only the quotient \verb?q? is output.
\item \verb?void RI_div_remainder(RecInt<k>& r, const RecInt<k>& a , const RecInt<k>& b)?: only the remainder \verb?r? is output.
\item \verb?void RI_square(RecInt<k>& ah, RecInt<k>& al, const RecInt<k>& b)?: $ah*2^{2^k}+al=b^2$.
\item \verb?void RI_gcd(RecInt<k>& g, const RecInt<k>& a, const RecInt<k>& b)?: \verb?g? is set to the Gcd of \verb?a? and \verb?b?.
\item \verb?void RI_ext_gcd(RecInt<k>& g, bool& su, RecInt<k>& u, bool& sv,?\\ \verb?RecInt<k>& v, const RecInt<k>& a, const RecInt<k>& b)?: \verb?g? is set to the Gcd of \verb?a? and \verb?b? with Bezout coefficients \verb?u? and \verb?v? respectively with sign \verb?su? and \verb?sv? (\verb?su==0? means that \verb?u? is negative).\\
\end{itemize}

Addition and subtraction functions are provided with extended functions having suffixes \verb?_in?, \verb?_nc? or combined \verb?_nc_in?.\\

The \verb?_in? suffix means that the operation is made in place:\\
\verb?void RI_add_in(limb& l, RecInt<k>& a, const RecInt<k>& b)?: $a\leftarrow a+b$ and \verb?l? is the carry.\\

The \verb?_nc? suffix means that the carry (or borrow) is not output:\\
\verb?void RI_sub_nc(RecInt<k>& a, const RecInt<k>& b, const RecInt<k>& c)?: $a\leftarrow b-c$.

\subsection{Extending the word size}

The following functions return results modulo $2^{2^k}$:

\begin{itemize}
  \item \verb?void RI_add_nc(RecInt<k>& a, const RecInt<k>& b, const RecInt<k>& c)? returns \verb?a? such that \verb?a=b+c? modulo $2^{2^k}$.
  \item \verb?void RI_sub_nc(RecInt<k>& a, const RecInt<k>& b, const RecInt<k>& c)? returns \verb?a? such that \verb?a=b-c? modulo $2^{2^k}$.
  \item \verb?void RI_mul(RecInt<k>& a, const RecInt<k>& b, const RecInt<k>& c)? returns \verb?a? such that \verb?a=b*c? modulo $2^{2^k}$.
\end{itemize}

\subsection{Modular operations}
The PALOALTO Library allows manipulation of modular integers as well.

\subsubsection{Modular operations on RecInt}
Our library provides modular operations on classical \verb?RecInt?. The user must specify the module $n$ at each call of any operation. At the beginning of all functions, the input parameters are reduced modulo $n$. The outputs are also computed modulo $n$. 

\begin{itemize}
\item \verb?void RI_reduction(RecInt<k>& a, const RecInt<k>& b, const RecInt<k>& n)?: $a\leftarrow b\text{ mod }n$.
\item \verb?void RI_random_mod(RecInt<k>& a, const RecInt<k>& n)?: \verb?a? is set to a random value within the range $0..n-1$.
\item \verb?void RI_neg_mod(RecInt<k>& a, const RecInt<k>& b, const RecInt<k>& n)?: $a\leftarrow -b\text{ mod }n$.
\item \verb?void RI_add_mod(RecInt<k>& a, const RecInt<k>& b, const RecInt<k>& c,?\\ \verb?const RecInt<k>& n)?: $a\leftarrow b+c\text{ mod }n$.
\item \verb?void RI_sub_mod(RecInt<k>& a, const RecInt<k>& b, const RecInt<k>& c,?\\ \verb?const RecInt<k>& n)?: $a\leftarrow b-c\text{ mod }n$.
\item \verb?void RI_mul_mod(RecInt<k>& a, const RecInt<k>& b, const RecInt<k>& c,?\\ \verb?const RecInt<k>& n)?: $a\leftarrow b*c\text{ mod }n$.
\item \verb?void RI_mul_mod(RecInt<k>& a, const RecInt<k>& b, const limb& c,?\\ \verb?const RecInt<k>& n)?: $a\leftarrow b*c\text{ mod }n$.
\item \verb?void RI_square_mod(RecInt<k>& a, const RecInt<k>& b, const RecInt<k>& n)?: $a\leftarrow b^2\text{ mod }n$.
\item \verb?void RI_exp_mod(RecInt<k>& a, const RecInt<k>& b, const RecInt<k>& c,?\\ \verb?const RecInt<k>& n)?: $a\leftarrow b^c\text{ mod }n$.
\item \verb?void RI_exp_mod(RecInt<k>& a, const RecInt<k>& b, const limb& c,?\\ \verb?const RecInt<k>& n)?: $a\leftarrow b^c\text{ mod }n$.
\item \verb?void RI_inv_mod(RecInt<k>& a, const RecInt<k>& b, const RecInt<k>& n)?: $a\leftarrow b^{-1}\text{ mod }n$.
\item \verb?void RI_div_mod(RecInt<k>& a, const RecInt<k>& b, const RecInt<k>& c,?\\ \verb?const RecInt<k>& n)?: $a\leftarrow b*c^{-1}\text{ mod }n$ if \verb?c? is invertible modulo \verb?n?.
\item \verb?bool RI_is_quadratic_residue(const RecInt<k>& a, const RecInt<k>& n)? returns \verb?true? if and only if \verb?a? is a quadratic residue modulo \verb?n?.
\item \verb?void RI_square_root_mod(RecInt<k>& a, const RecInt<k>& b, const RecInt<k>& n)?: \verb?a? is such that $a^2= b\text{ mod }n$.

\end{itemize}

The \verb?neg?, \verb?add? and \verb?sub? functions are extended with the suffix \verb?_in? (in place operation) as for the classic operations. Note that the \verb?_nc? suffix does not make any sense in this situation.

\subsubsection{RecIntMod type}

A special type is provided for modular operations. \verb?RecIntMod<>? is basically a \verb?RecInt<>? provided with a module $p$ declared as a \verb?static RecInt<>? of the same size. Furthermore we guarantee that such elements are always reduced modulo $p$. Actually we guarantee that all inputs and outputs of functions manipulating \verb?RecIntMod<>? are reduced modulo $p$ (that is to say within the range $0..p-1$).

\begin{verbatim}
template <size_t k> struct RecIntMod {
  typedef RecInt<k> RecIntk;

  static RecIntk p;
  RecIntk Value;
};
\end{verbatim}

Two functions allowing the conversion from \verb?RecIntMod<>? to \verb?RecInt<>? and vice versa are available: \verb?Convert_to_RecInt(RecInt<k>&, RecIntMod<k>)? \\ and \verb?Convert_to_RecIntMod(RecIntMod<k>&, RecInt<k>)?. \\

\subsubsection{Modular operations on RecIntMod}
The same operations as in last section are applicable to \verb?RecIntMod<>? integers. However the module \verb?p? must be initialized before any arithmetic operation with the following function: \\
\verb?void RI_init_module(const RecInt<k>& p)?\\

Since the module is declared as \verb?static?, one has to initialize it only once before any computation. If done so, we guarantee that any \verb?RecIntMod<>? integer will be always reduced modulo \verb?p? throughout the program. Note that the user is allowed to change the module a posteriori, but the reduction modulo the new module will not be guaranteed anymore. \\

The user can get the module back by using:\\
\verb?void RI_get_module(RecInt<k>& p)?\\

Once the module has been initialized, the user can use the operations presented in \emph{Modular operations on RecInt} section, since they are overloaded for a use with \verb?RecIntMod?.\\

\section{C++ library Performances}

In order to evaluate the performance of the PALOALTO prototype, we used GMP's assembly routine for double machine word arithmetic (e.g. \verb?umul_ppmm? defined in the GMP \cite{GMP} file \verb?longlong.h?, multiplying two integers and generating their two-word product). \\
Using these assembly routines and the recursive data structure detailed above, we were able to get the performance of the following tables for fixed precision operations with recursive data structures.\\
For these performance comparisons we use the GMPbench suite \cite{GMPb}.

\begin{figure}[H]
  \begin{center}
    \includegraphics[width=300px]{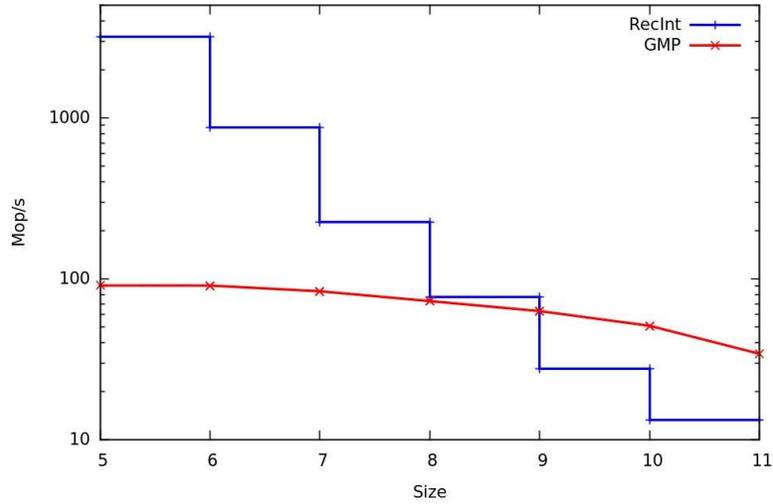} \\
    \ \\
    \caption{Fixed precision addition with RecInt versus GMP-5.0.1, gcc 4.4.0, Xeon X5482, 3.2GHz, in millions of arithmetic operations per second.}
  \end{center}
\end{figure}

\begin{figure}[H]
  \begin{center}
    \includegraphics[width=300px]{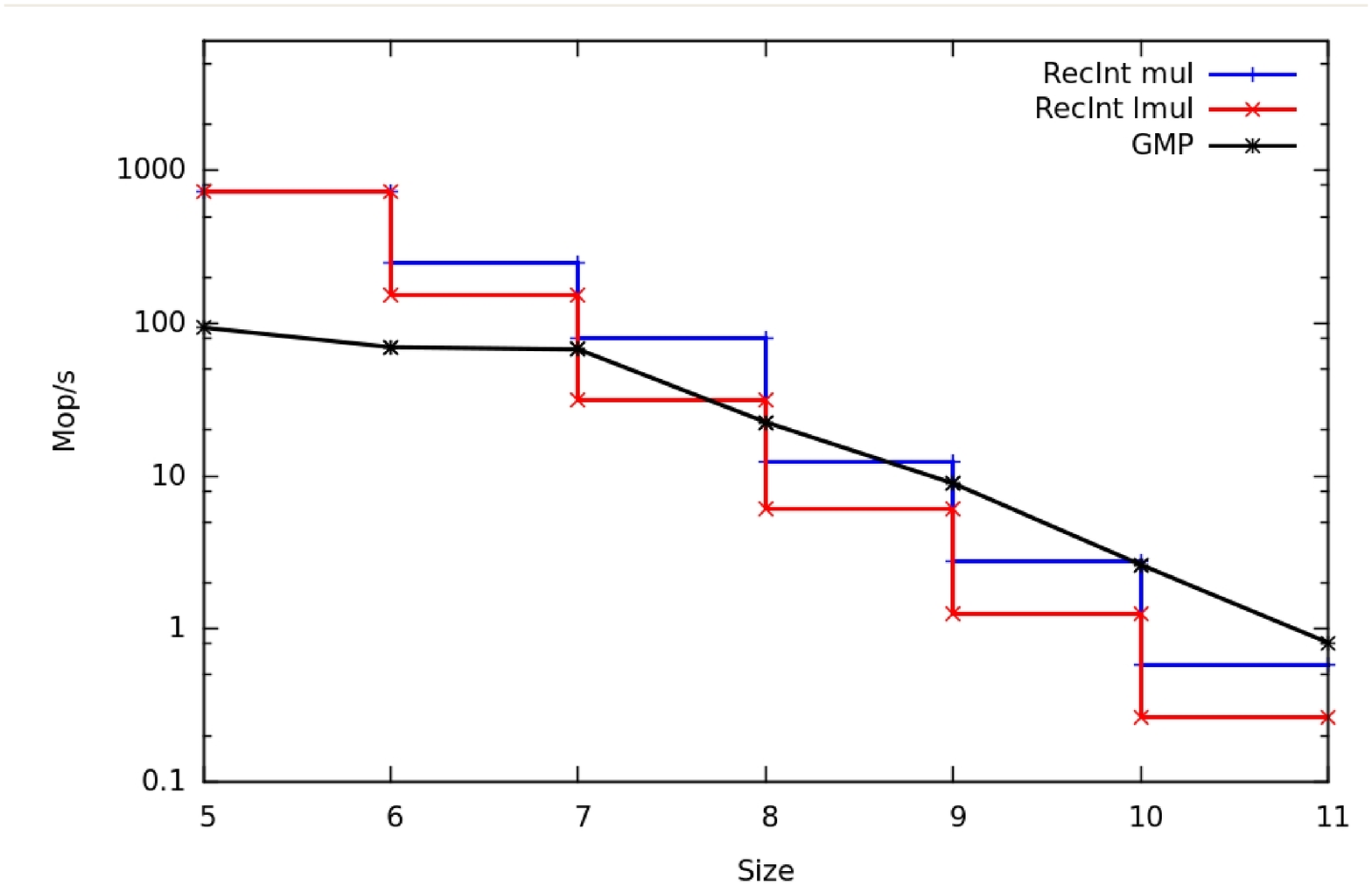} \\
    \ \\
    \caption{Fixed precision complete and truncated multiplications with RecInt versus GMP-5.0.1, gcc 4.4.0, Xeon X5482, 3.2GHz, in millions of arithmetic operations per second.}
  \end{center}
\end{figure}

\begin{figure}[H]
  \begin{center}
    \includegraphics[width=300px]{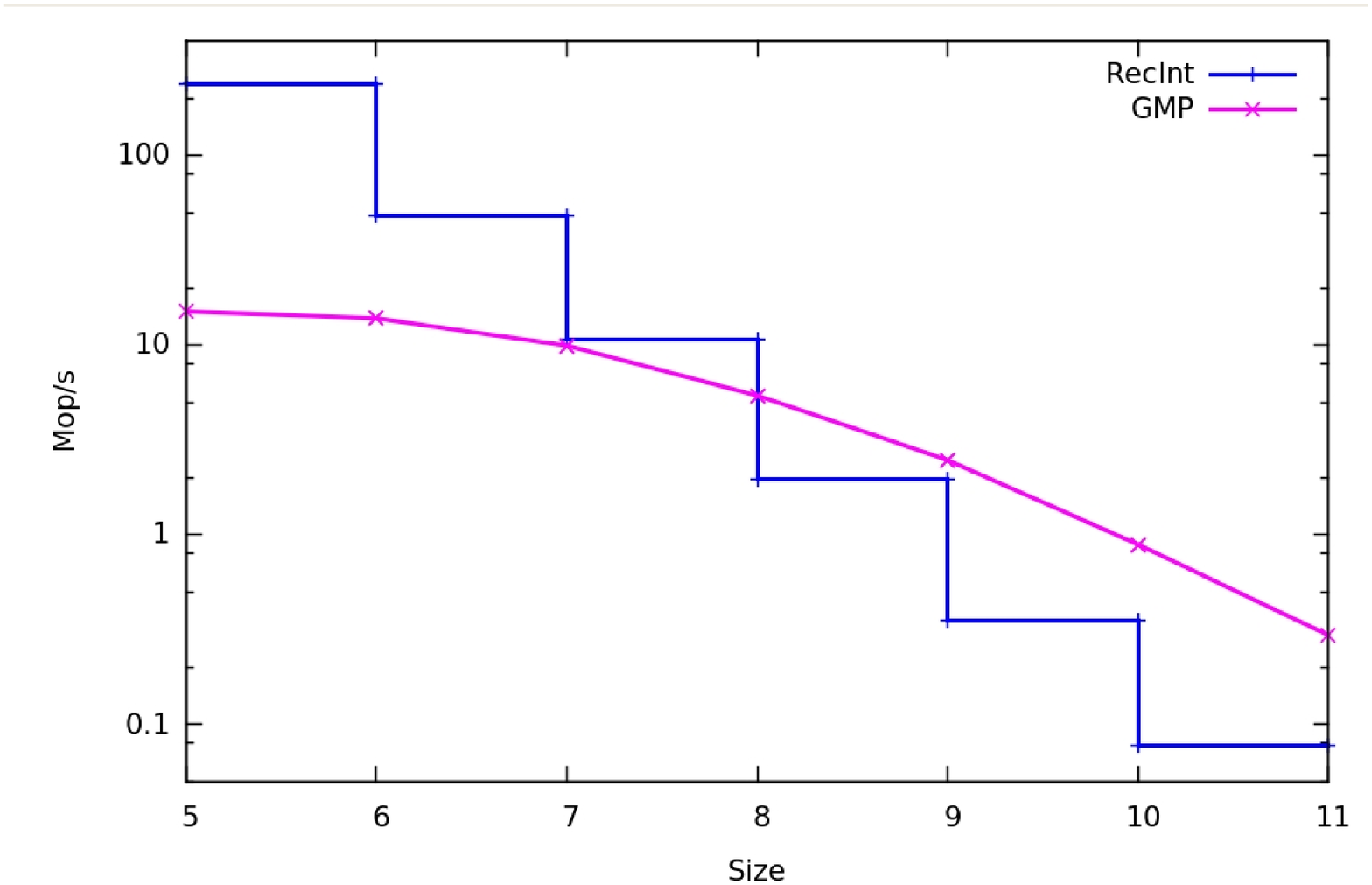} \\
    \ \\
    \caption{Fixed precision modular multiplication with RecInt versus GMP-5.0.1, gcc 4.4.0, Xeon X5482, 3.2GHz, in millions of arithmetic operations per second.}
  \end{center}
\end{figure}

\begin{figure}[H]
  \begin{center}
    \includegraphics[width=300px]{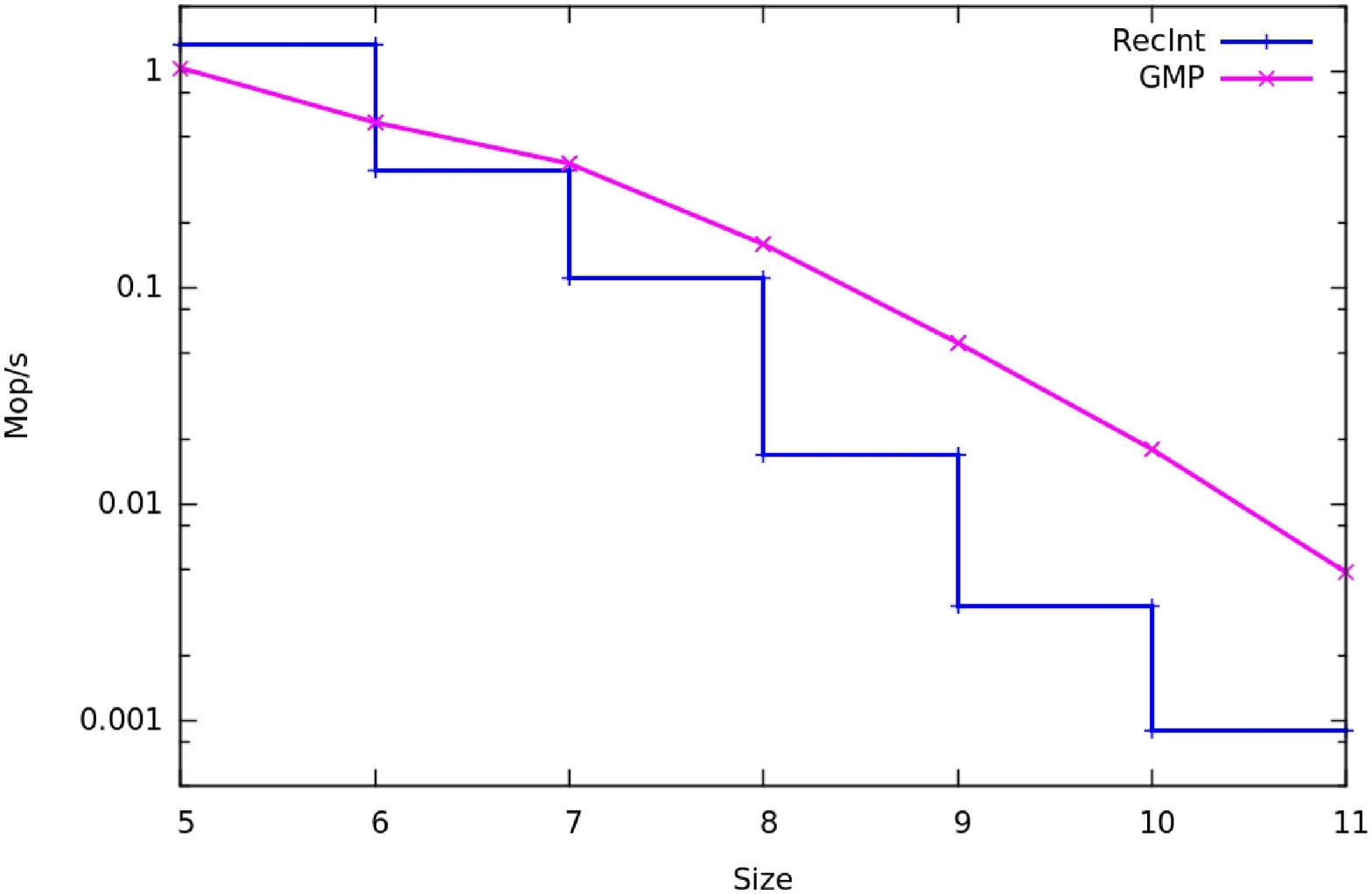} \\
    \ \\
    \caption{Fixed precision modular exponentiation with RecInt versus GMP-5.0.1, gcc 4.4.0, Xeon X5482, 3.2GHz, in millions of arithmetic operations per second.}
  \end{center}
\end{figure}

The step shape of RecInt curves is explained by the fact that we use a \verb?RecInt<?$k$\verb?>? for all integers with size within the range $2^{k-1}-1..2^k$. \\

Results obtained with RecInt are comparable to those with GMP. However, RecInt appears to be more efficient for small fixed precision.

\section{Towards FPGA implementation}

We present here the first attempts towards an implementation on a real FPGA. \\
Inside SHIVA project, we need to provide basic arithmetic modules in order to be used in a RSA or Elliptic curve based encryption scheme. In order to build these modules, since our C++ library is already written, we chose to use a dedicated software transforming C++ source into VHDL called GAUT \cite{GAUT}. The creation of VHDL program can be split in the following steps:
\begin{itemize}
\item Compilation of C++ source and creation of the corresponding graph.
\item Compilation of the library containing the needed operations.
\item Synthesizing of the VHDL program and estimation of performances. \\
\end{itemize}

Here are some simulations of a modular exponentiation on a Virtex 5. We made the output flow vary in order to check the effect on the required size on the FPGA.

\begin{figure}[H]
  \begin{center}
    \includegraphics[width=300px]{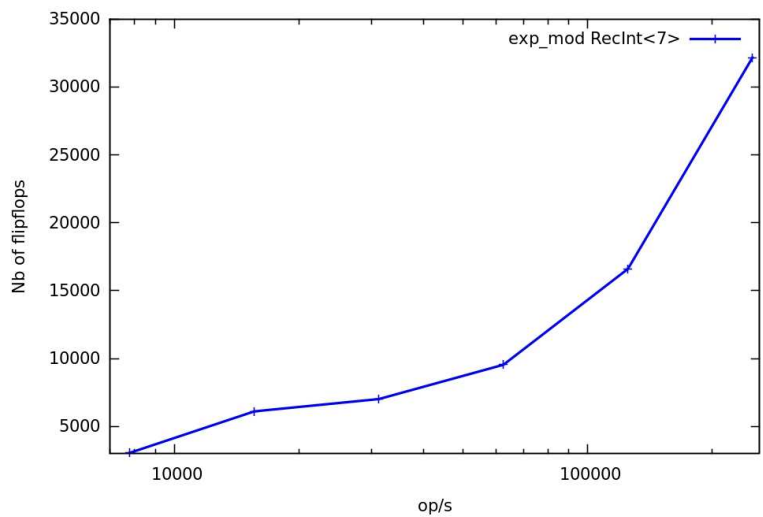} \\
    \caption{128 bits words.}
  \end{center}
\end{figure}

\begin{figure}[H]
  \begin{center}
    \includegraphics[width=300px]{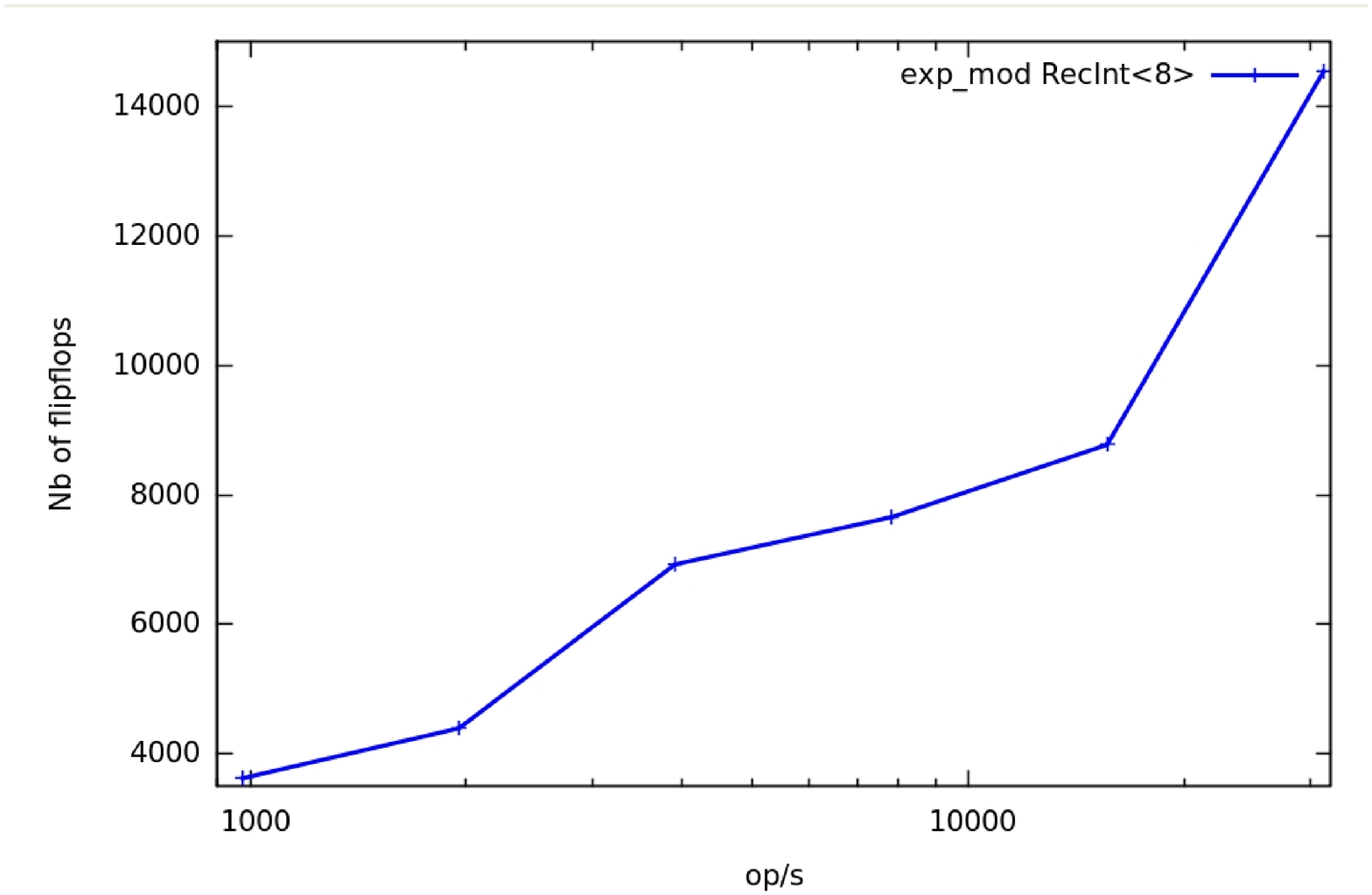} \\
    \caption{256 bits words.}
  \end{center}
\end{figure}

\begin{figure}[H]
  \begin{center}
    \includegraphics[width=300px]{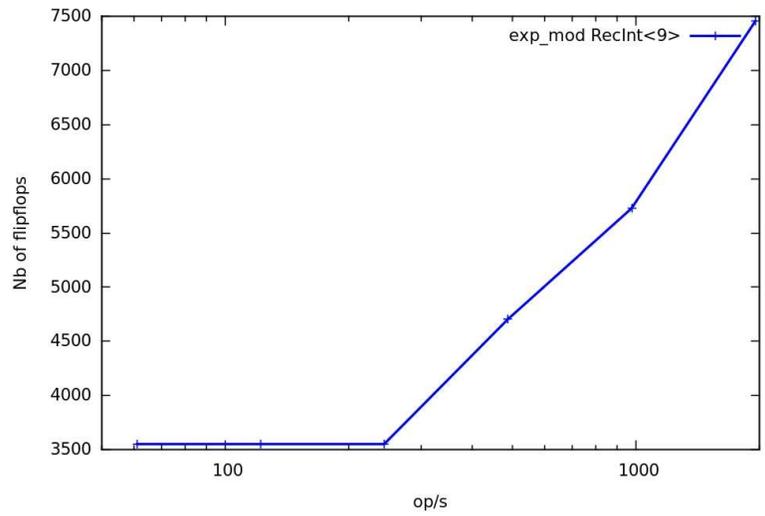} \\
    \caption{512 bits words.}
  \end{center}
\end{figure}

We notice that required size can be significantly reduced if we accept a lower output flow. \\

These results have been obtained without significant modifications on the C++ source. Thus they are not optimal but rather promising. \\

Further work will consist in optimizing C++ source in order to make it more adapted to VHDL synthesizing.

\end{document}